\documentclass[runningheads]{llncs}
\usepackage[T1]{fontenc}
\usepackage{graphicx}
\usepackage[table]{xcolor}
\usepackage{tikz}
\usepackage{makecell}
\usepackage{booktabs}
\usepackage{comment}
\usepackage{subcaption}
\usepackage{url}
\usepackage[hidelinks]{hyperref}

\usepackage[most]{tcolorbox}
\tcbset{
  keyfindingbox/.style={
    colback=gray!10, colframe=black!70,
    boxrule=0.5pt, arc=2pt, outer arc=2pt,
    left=4pt, right=4pt, top=2pt, bottom=0pt,
    fonttitle=\bfseries 
  }
}

\newlength{\compactBeforeTable}
\newlength{\compactAfterTable}
\newlength{\compactBeforeFigure}
\newlength{\compactAfterFigure}
\newlength{\compactBeforeSection}
\newlength{\compactAfterSection}
\newlength{\compactBeforeSubsection}
\newlength{\compactAfterSubsection}
\newlength{\compactBeforeParagraph}
\newlength{\compactAfterParagraph}
\newif\iftechrep

\techrepfalse

\newcommand{\appref}[1]{%
\iftechrep
Appendix~\ref{#1}%
\else
the technical report of the paper~\cite{techrep}%
\fi
}

\begin{document}
\title{Layered Ego Networks in Email Communication: From Enron to the Jmail Archive}
\titlerunning{Layered Ego Networks in Email Communication}
%
\author{Francesco Di Cursi\inst{1,2}\orcidID{0009-0005-9782-478X} \and
Chiara Boldrini\inst{1}\orcidID{0000-0001-5080-8110} \and
Marco Conti\inst{1}\orcidID{0000-0003-4097-4064} \and
Andrea Passarella\inst{1}\orcidID{0000-0002-1694-612X}}
\authorrunning{F. Di Cursi et al.}
%
\institute{Institute of Informatics and Telematics, Pisa  56124, IT \\
\email{\{francesco.dicursi,chiara.boldrini,marco.conti,andrea.passarella\}@iit.cnr.it} \and
University of Pisa, Pisa, 56127, IT\\
}
\maketitle              
\begin{abstract}
Email archives offer a rare view of social relationships through repeated communication, but it remains unclear how well classical ego network layering applies to digital interaction data. This paper compares two public email archives with sharply contrasting structures: Enron, a workplace corpus involving around 150 users, and Jmail, a single-ego archive centered on an exceptionally active focal actor whose communication volume is more than twenty times higher than the average Enron user. We ask, in each case, whether Dunbar-like layered organization is recoverable from email communication frequency and how it should be interpreted.
For Jmail, we show that extreme communication intensity causes standard layering methods (whether clustering-based or threshold-based) to break down. Jmail is not a broad communication environment with many occasional contacts, but a selective pool of high-interest alters operating on a much higher frequency scale than ordinary email. Once the Dunbar frequency ladder is anchored to the empirical support-clique boundary, a clearer layered structure emerges. Reciprocity analysis confirms that the recovered layers reflect genuine bidirectional relationships rather than artifacts of the focal actor's outgoing activity.
Enron serves as a workplace benchmark that grounds the comparison: its ego networks partially reproduce Dunbar-like organization, with stable inner circles and an outermost recovered layer corresponding to Dunbar's affinity group ($\sim50$), confirming that layered structure is recoverable from ordinary organizational email.
Overall, the findings show that Dunbar-like organization can be meaningfully studied in email archives, but that selective high-frequency archives require frequency normalization before the layered structure becomes interpretable.



\keywords{Email Communication  \and Ego networks \and Dunbar's model \and Temporal dynamics \and Enron \and Jmail}
\end{abstract}
%
%
\vspace{-1cm}
\section{Introduction}
\vspace{\compactAfterSection}

Email is one of the oldest and most persistent forms of digital communication, yet it remains among the least studied from a social network perspective. Unlike platforms designed for public sharing, email encodes social relationships through repeated private interaction: ties are not declared but enacted, message by message, over months and years. This makes email logs a uniquely faithful
record of real communicative investment. Yet the same properties that make email valuable also make it difficult to study. Privacy constraints, limited accessibility, and the sensitivity of personal correspondence keep public datasets scarce, with only a handful of widely used resources available for research~\cite{klimt2004enron,oard2015avocado,leskovec2014snap}.

This scarcity has a concrete cost for ego network research~\cite{dunbar2015structure}. Ego networks provide a user-centered view of social structure, placing a focal individual (the ego) at the center and representing their contacts (the alters) as a surrounding personal network. Rather than studying global network topology, ego network analysis asks how communication is distributed across an individual's contacts and how stronger and weaker ties organize within that personal sphere. A central finding of this literature, most prominently associated with the work of Robin Dunbar, is that personal networks are not flat collections of contacts but are organized into discrete layers of increasing size and decreasing interaction intensity~\cite{dunbar1998social,dunbar2015structure}. In Dunbar's model, these layers follow a canonical structure: a support clique of around 5 alters contacted at least weekly, a sympathy group of around 15 contacted at least monthly, an affinity group of around 50 contacted at least every six months, and an active network of around 150 contacted at least yearly. Successive layers scale by a ratio of approximately 3, and interaction frequency drops correspondingly as one moves outward. More recently, an innermost layer of just 1--2 alters has been identified, representing the ego's closest and most frequently contacted ties~\cite{dunbar2015structure}.  Crucially, this structure is thought to reflect cognitive and time constraints on relationship maintenance rather than platform-specific affordances, which is why it has been proposed as a universal feature of human social networks. Prior work has confirmed that these patterns are not confined to offline relationships but re-emerge in digital communication traces, including mobile-phone calls and online social platforms~\cite{arnaboldi2012analysis,arnaboldi2015online,mac2016calling}, and have recently been observed in decentralized social media as well~\cite{di2024herd}. 
Yet email remains comparatively underexplored as a setting for testing Dunbar-like ego network organization. It is therefore still unclear whether ego networks extracted from email communication reproduce layered organization in both circle size and contact frequency.

This paper addresses that gap directly. We analyse two public email archives with sharply contrasting structures and ask, in each case, whether layered ego network organization is recoverable and how it should be interpreted. The first archive is Enron, a large organizational corpus involving around 150 users whose complete mailboxes were made public during a federal investigation. Despite being one of the most studied datasets in email network research, its ego network structure has not been examined through the lens of Dunbar-style layering. We use it as a workplace benchmark: a setting in which communication is recurrent and role-structured, and in which Dunbar-like layering, if present, should be most clearly visible. Our analysis shows that it is: Enron users can be organized into layered personal networks with progressively larger and less frequently contacted circles, providing the first explicit evaluation of Dunbar-like ego network structure in workplace email.

We then use this baseline to study Jmail, a rare and structurally unusual archive. Jmail is centered on a single, highly active focal actor whose communication record spans approximately ten years and involves thousands of distinct contacts, at a volume more than twenty times higher than the average Enron user. Public email datasets are rare; large archives centered on a single socially salient individual and available for research are rarer still. This makes Jmail a valuable but non-standard test case: it does not represent ordinary email communication, but a selective, partially redacted record of an exceptionally intense ego-centered communication space. Applying standard ego network methods to such an archive is non-trivial, and understanding where and why they break down (and how they can be recovered) is of both methodological and empirical interest.

Against this background, this work makes three contributions. First, it characterizes Jmail as a rare and structurally unusual ego-centered archive in which extreme communication intensity causes the standard methods for obtaining Dunbar's circles (whether clustering-based or threshold-based) to break down, and shows that rescaling contact-frequency thresholds to the archive's communication intensity recovers a meaningful layered structure, confirmed as genuinely bidirectional through reciprocity analysis. Second, it evaluates Dunbar-like ego network layers in email communication, using Enron as a workplace benchmark that grounds the comparison and makes the unusual properties of Jmail interpretable. Third, it shows that Dunbar's model is robust across these contrasting regimes: layered organization emerges in ordinary workplace email and can be recovered in high-volume ego-centered archives alike, provided that frequency thresholds are adapted to the communication scale of the dataset.

The remainder of the paper is organized as follows. Section~\ref{sec:prev_works} reviews ego network studies in offline, online, and email-based social environments. Section~\ref{sec:methodology} describes the email archives and the construction of layered ego networks. Section~\ref{sec:results} presents the empirical analysis, including the Enron ego network analysis, the emergence of Dunbar-like structures in email, the contrasting organization of Jmail, and the comparison between data-driven, direct Dunbar-based, and rescaled Dunbar-style groupings. Section~\ref{sec:conclusion} concludes the paper.

\vspace{\compactBeforeSection}
\section{Related works}
\label{sec:prev_works}
\vspace{\compactAfterSection}

This work connects ego network research, where interaction frequency distinguishes stronger and weaker ties around an ego, with email-based social network analysis, where communication logs reveal temporal rhythms, roles, communities, and organizational structure.
\vspace{\compactBeforeParagraph}
\paragraph{Ego networks in offline and online environments.}

ego network research has shown that personal social networks are not homogeneous collections of contacts, but tend to be organized into layers of increasing size and decreasing interaction intensity. This regularity is usually associated with Dunbar's model, where alters are arranged around the ego in concentric circles, from a small support core to a broader active network \cite{dunbar1998social,dunbar2015structure}. Table~\ref{tab:egonet_offline} summarizes the offline reference sizes and contact-frequency thresholds provided by Dunbar, which we will use later in the paper.
Empirical studies in online social networks have shown that digital platforms can reproduce comparable ego network constraints, despite differences in interaction cost and medium \cite{arnaboldi2012analysis,arnaboldi2015online}. Interestingly, these online studies have also revealed an innermost layer of just 1--2 alters, representing the ego's closest and most frequently contacted ties, which is not part of the canonical offline model but emerges consistently in digital interaction data~\cite{dunbar2015structure}. Similar layered patterns have also been observed in mobile-phone communication, where call frequency is used to cluster alters into ego network layers comparable to Dunbar's predicted circle sizes \cite{mac2016calling}. More recent ego-centered analyses of decentralized social platforms have extended this perspective to newer forms of online interaction \cite{di2024herd}.
These works provide the structural reference for the present study. They show that layered ego networks can be recovered from behavioral traces, and that interaction frequency is a meaningful proxy for tie strength. However, most evidence comes from offline relationships, online social platforms, or mobile-phone communication. Email remains comparatively less explored from this perspective, and prior work has not systematically asked whether email-derived ego networks recover Dunbar-like layers in terms of both circle size and contact frequency.

\begin{table}[t!]
\tiny
\centering
\caption{Reference properties of offline ego network layers.}
\label{tab:egonet_offline}
\resizebox{\linewidth}{!}{
\begin{tabular}{ c  c  c  c  c  c }
\toprule
\makecell{\textbf{Layer}\\\textbf{number}} &
\makecell{\textbf{Layer}\\\textbf{name}} &
\textbf{Description} &
\textbf{Alters} &
\makecell{\textbf{Scaling ratio}\\\textbf{(alters)}} &
\makecell{\textbf{Contact}\\\textbf{frequency}} \\
\midrule
1 & Support clique & close family & 5 & & once every week \\
2 & Sympathy group & close friends & 15 & $\sim 3$ & once a month \\
3 & Affinity group & \makecell{friends/extended family/\\colleagues} & 50 & & once every 6 months \\
4 & Active network & meaningful relationships & 150 & & once a year \\
\bottomrule \vspace{-20pt}
\end{tabular}
}
\end{table}

\vspace{-15pt}
\paragraph{Emails and social network analysis.}

Email has long been used to study social structure, organizational communication, and temporal interaction dynamics. Early work showed that email logs can reveal communities of practice within organizations, using message exchange as a proxy for information flow and social proximity~\cite{tyler2005mail}. Other studies emphasized temporal organization: Barabási showed that human actions, including email activity, are bursty and heavy-tailed rather than Poisson-like~\cite{barabasi2005origin}, while Eckmann, Moses, and Sergi showed that email traffic contains coherent dialogue structures based on repeated and synchronized exchanges~\cite{eckmann2004entropy}. Large-scale messaging studies also found strong daily and weekly rhythms, burstiness, and uneven activity across users and ties~\cite{golder2007rhythms}. Together, these results indicate that email and messaging networks are heterogeneous both in time and across relationships.
A second line of work has focused on the structure and evolution of email networks. Kossinets and Watts analysed a university email network of 43,553 individuals over one academic year, showing that tie formation is shaped by organizational affiliation, similarity, and triadic closure~\cite{kossinets2006empirical}. In the Enron corpus, Diesner and Carley extracted communication networks from the email archive and showed that the network became denser, more centralized, and more connected during the organizational crisis~\cite{diesner2005exploration}. McCallum et al. combined email structure and content to discover topics and roles in Enron and academic email, showing that communication patterns are shaped by sender--recipient relations and organizational roles~\cite{mccallum2007topic}. These studies establish Enron as one of the main benchmarks for email-based social network analysis, but they mainly focus on global network structure, community organization, roles, crisis dynamics, or topic-aware interaction patterns.

The present work differs from this literature by using Enron as a benchmark for ego network layer extraction in workplace email and Jmail as an extreme-frequency ego-centered archive. To the best of our knowledge, prior email-network studies have documented skewed communication, bursty timing, strong and weak tie regimes, and dense organizational cores, but have not translated these patterns into a longitudinal Dunbar-style ego network analysis. Our contribution is therefore a comparison between a conventional workplace corpus and a selective high-contact archive, aimed at testing when Dunbar-like structure is visible, distorted, or recoverable through frequency scaling.
\vspace{-0.5cm}

\section{Methodology}
\label{sec:methodology}
\vspace{\compactBeforeSubsection}
\subsection{Email Archives}
\vspace{\compactAfterSubsection}


We ground our analysis in two public email archives that differ sharply in scope and structure, allowing us to test layered ego network methods in a conventional organizational setting before applying them to a more unusual ego-centered archive.

\paragraph{Enron email corpus.} One of the most widely used public datasets for the study of email communication. The version used in this work is the Carnegie Mellon University CALO Project release.\footnote{\url{https://www.cs.cmu.edu/~enron/}} The corpus contains approximately half a million messages from about 150 users, mostly members of Enron senior management. The data were originally made public by the Federal Energy Regulatory Commission during its investigation of Enron. In the present study, Enron serves as a large-scale organizational benchmark, representative of a conventional workplace communication setting. 
Table~\ref{tab:summary_datasets} reports statistics at two stages of the processing pipeline. \textit{Nodes} and \textit{Links} describe the global communication graph. \textit{Unique emails} counts original messages prior to recipient expansion, whereas \textit{Interactions} counts exploded sender–recipient pairs; consequently, a single email sent to multiple recipients contributes multiple interactions. The ego-level columns (\textit{Egos}, \textit{Alters}, and \textit{Alter-Egos}) are defined only for the selected scope, which is restricted to the 158 users with complete mailboxes. This restriction is necessary because only these users support full ego network reconstruction. Within this subset, the dataset contains 91,990 unique emails, 115,384 interactions, 11,811 alters, and 151 egos that also appear as alters in other ego networks.


\begin{table*}[!t]
\tiny
\centering
\caption{Summary statistics of the Enron and Jmail datasets. Note that both graphs are directed. Ego-Alters are egos who also appear as alters in other ego networks.}
\label{tab:summary_datasets}
\resizebox{\textwidth}{!}{%
\begin{tabular}{lrrrrrrr}
\toprule
 & \textbf{Nodes} & \textbf{Links} & \textbf{Uniq. emails} & \textbf{Interactions} & \textbf{Egos} & \textbf{Alters} & \textbf{Ego-Alters} \\
\cmidrule(r){1-5}\cmidrule(l){6-8}
Enron (raw)      & 35,848 & 104,449 & 495,554   & 730,852   & --- & --- & ---  \\
Enron (selected) & 11,818 & 22,249  & 91,990    & 115,384   & 158    & 11,811 & 151    \\
\cmidrule(r){1-5}\cmidrule(l){6-8}
Jmail (raw)      & 82,968 & 185,539 & 1,148,096 & 1,293,703 & --- & --- & --- \\
Jmail (selected) & 4,299  & 4,299   & 287,630   & 308,515   & 1      & 4,299  & ---    \\
\bottomrule
\end{tabular}%
} \vspace{-10pt}
\end{table*}



\vspace{\compactBeforeParagraph}
\paragraph{Jmail.}

The second dataset is based on a publicly released archival email collection.\footnote{{\tiny We study Jmail as a public archival dataset and focus only on aggregate communication structure. The analysis does not infer private attributes of individuals or evaluate the content of messages.}} The collection documents the email activity of a high-profile public figure whose correspondence was released as part of a legal investigation. In its original distribution, the material consists primarily of redacted PDF files, which makes direct computational analysis non-trivial. We ultimately relied on the structured email representation provided through Jmail\footnote{{\tiny\url{https://jmail.world}. The Jmail file used in the present work is \texttt{jmail\_data/v1/emails.parquet}}}, which has become a de facto reference format for accessing this collection and also functions as an API-like access layer allowing direct retrieval of the public data. Unlike Enron, Jmail is not a broad organizational corpus: it is centered on one main actor whose core activity spans approximately ten years, offering a selective and partially redacted view of a highly active ego-centered communication space.

Table~\ref{tab:summary_datasets} reports the statistics about the Jmail dataset. The ego-level columns are defined only for the selected scope, which is restricted to the single main actor around whom the archive is organized. Within this scope, the dataset contains 287,630 unique emails, 308,515 interactions, and 4,299 alters. Additional details on preprocessing, identity consolidation, and filtering are reported in~\appref{app:jmail_preprocessing}.
Table~\ref{tab:main_actors_activity_intensity} provides additional information about Enron and Jmail at the ego level. Although Jmail contains only one focal ego, this ego is observed over a much longer lifespan and produces substantially higher activity than the average Enron main actor. In particular, the Jmail main actor produces 21.79 messages per day, compared with 0.93 messages per day for Enron main actors on average, i.e., more than twenty times higher activity. This confirms that Jmail is not simply a smaller ego-centered version of Enron, but an unusually intense single-ego archive.

\begin{table}[t]
\tiny
\centering
\caption{Activity intensity for Enron and Jmail main actors.}
\label{tab:main_actors_activity_intensity}
\resizebox{0.9\textwidth}{!}{%
\begin{tabular}{llcccc}
\toprule
\textbf{Dataset} & \textbf{Scope} & \textbf{Egos} & \textbf{Period of activity [days]} & \textbf{Interactions/day} & \textbf{Messages/day} \\
\midrule
Enron & Main actors & 158 & 536.66 $\pm$ 251.63 & 1.20 $\pm$ 1.50 & 0.93 $\pm$ 1.14 \\
\midrule
Jmail & Main actor & 1 & 13192 & 23.37 & 21.79 \\
\bottomrule
\end{tabular}
} \vspace{-10pt}
\end{table}

\vspace{\compactBeforeSubsection}
\subsection{Ego Network Extraction}
\vspace{\compactAfterSubsection}

We provide here a concise description of the ego network extraction procedure, as it follows standard methods widely adopted in the related literature~\cite{dunbar2015structure}. 
For both archives, we represent email communication as a directed ego--alter network. Each email is converted into sender--recipient interactions: the sender is treated as the ego and each recipient as an alter. When a message has multiple recipients, it contributes one directed interaction for each sender--recipient pair. Before constructing the ego networks, we remove invalid records, missing identities, self-loops, duplicated interactions, and timestamps outside the selected observation windows.

Eligible egos are defined differently in the two datasets. In Enron, we consider only users represented by an individual mailbox folder, so that each ego has a comparable mailbox-level communication environment. In Jmail, the primary ego is the main actor around whom the archive is organized (by leveraging an apposite column indicating his identity).
%
The basic unit of analysis is the ego--alter dyad within a yearly window. For each ego and each alter, we count the number of observed directed interactions in that year. The procedure is applied to the core activity periods of the two datasets. For Enron, ego networks are extracted year by year for 2000 and 2001, which correspond to the main activity period retained in our analysis. For Jmail, yearly ego network snapshots are extracted from 2009 to 2018, capturing the core longitudinal activity of the focal actor and of the alters for whom ego network reconstruction is possible.


Layer extraction is performed only on active alters. Following Dunbar's model, an alter is considered active, i.e., occupying cognitive space in the ego's social network, only if the relationship meets a minimum level of sustained engagement: the ego--alter relation must span at least six months, involve at least two directed interactions, and include at least one interaction in the calendar year considered. This minimum threshold of roughly one contact per year reflects Dunbar's definition of a socially meaningful tie; alters that fall below it are retained for descriptive purposes but are not assigned to any social circle.
The goal of layer extraction is to identify the ego's social circles, i.e., the concentric groups of alters that correspond to Dunbar's layers. Following the standard approach in the literature~\cite{mac2016calling}, we recover these circles by clustering active alters according to their annual interaction frequency with the ego: alters contacted at similar rates are grouped together, and the resulting clusters are interpreted as social circles. 
Specifically, we use MeanShift clustering (a common choice in the related literature~\cite{arnaboldi2015online}), which has the practical advantage of not requiring the number of circles to be fixed in advance, allowing the empirical layering to emerge directly from the observed frequency distribution. Clustering is performed independently for each ego and year, and the resulting circles are ordered by decreasing mean annual frequency: circle~1 contains the highest-frequency alters, with subsequent circles containing progressively less frequently contacted alters. These data-driven circles provide the baseline empirical layering used in the comparison with Dunbar-based and rescaled Dunbar-style interpretations.

\begin{figure}[!h]
    \centering

    \begin{subfigure}[!h]{1\linewidth}
        \centering
        \includegraphics[trim=0 0 0 1.7cm,clip,width=\linewidth]{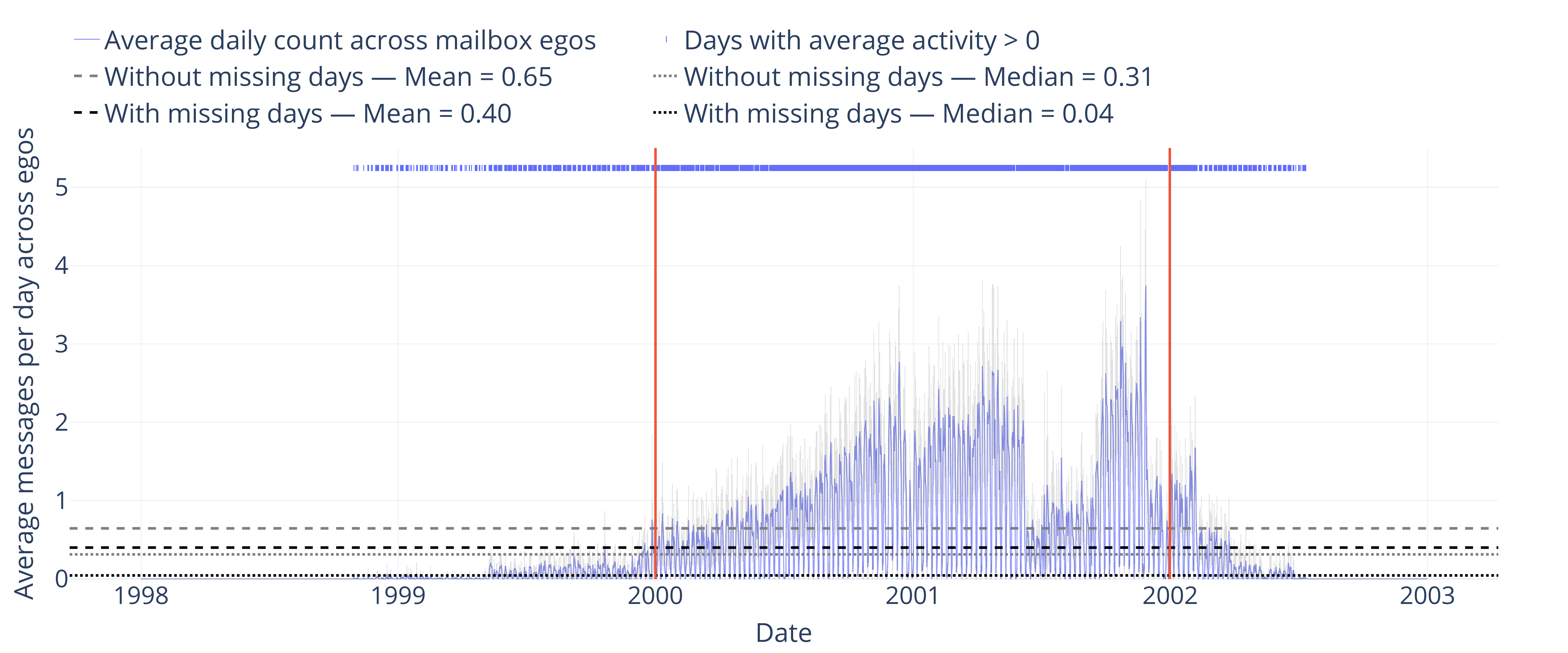}
        \caption{Enron --- Average daily activity}
        \label{fig:enron_daily}
    \end{subfigure}


    \begin{subfigure}[!h]{0.45\linewidth}
        \centering
        \includegraphics[trim=0 0 0 3cm,clip,width=\linewidth]{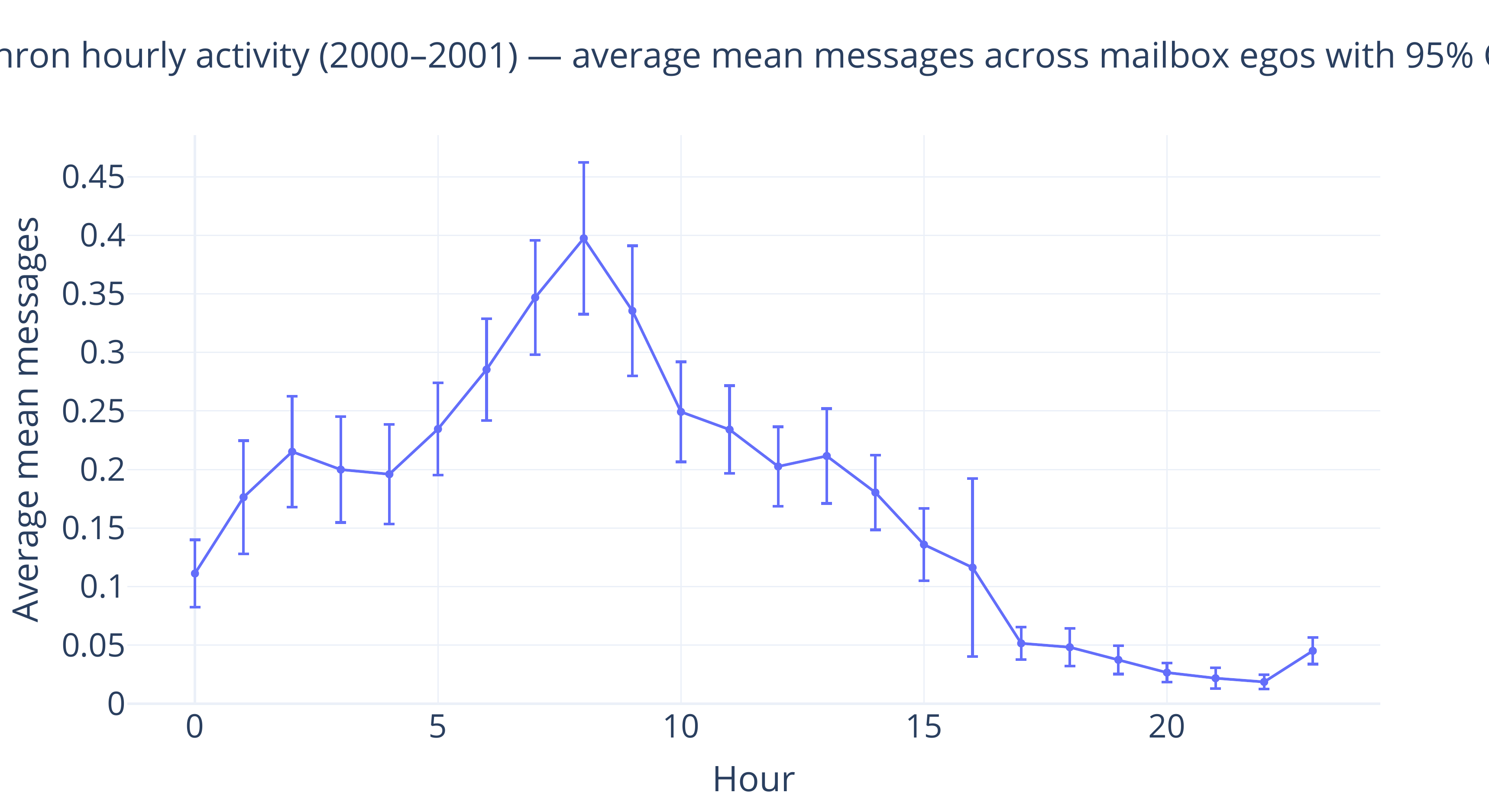}
        \caption{Hourly activity}
        \label{fig:enron_hourly}
    \end{subfigure}
    \hfill
    \begin{subfigure}[!h]{0.45\linewidth}
        \centering
        \includegraphics[trim=0 0 0 3cm,clip,width=\linewidth]{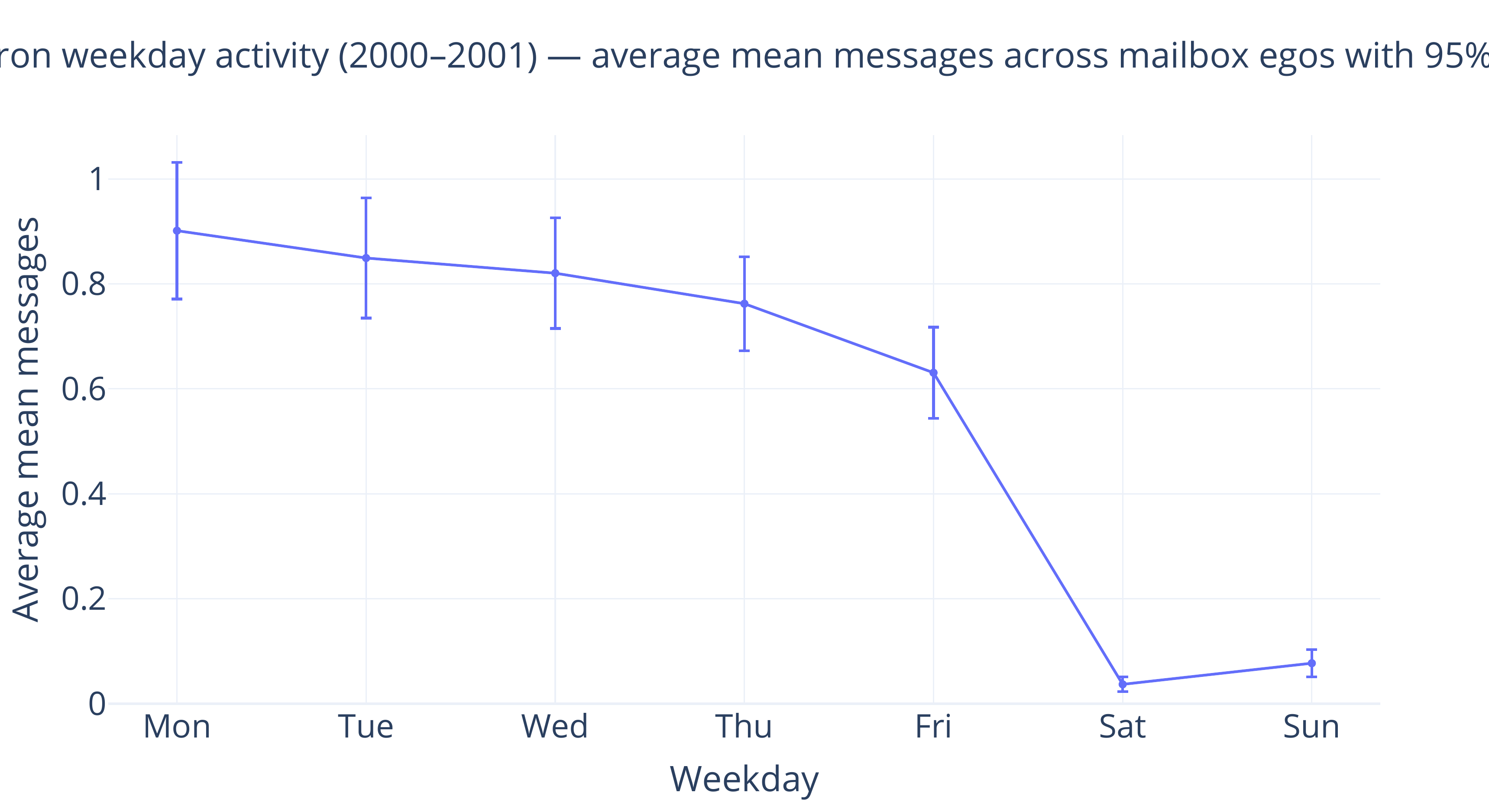}
        \caption{Weekday activity}
        \label{fig:enron_weekday}
    \end{subfigure}

    \caption{Enron (2000-2001) --- Temporal activity patterns of mailbox egos}
    \label{fig:enron_temporal_activity}
    \vspace{\compactAfterFigure}
\end{figure}

\vspace{-10pt}
\section{Results}
\vspace{-5pt}
\label{sec:results}


We report the results in two stages. We first use Enron as a conventional workplace-email benchmark to test whether layered ego network organization can be recovered from ordinary organizational communication (Sec.~\ref{sec:results_enron}). We then (Sec.~\ref{sec:results_jmail}) turn to Jmail, a substantially more extreme ego-centered archive, and show that standard Dunbar-style extraction methods break down under very high communication intensity. Finally, we show that rescaling the Dunbar frequency ladder restores an interpretable layered structure that remains reciprocally meaningful from both ego and alter perspectives.

\vspace{\compactBeforeSubsection}
\subsection{Enron as a Workplace Email Benchmark}
\label{sec:results_enron}
\vspace{\compactAfterSubsection}

\subsubsection{Temporal activity patterns.} Figure~\ref{fig:enron_temporal_activity} shows that Enron activity is concentrated in 2000 and 2001 and follows the expected rhythms of workplace communication: activity increases during working hours, declines in the evening, and drops substantially during weekends. These patterns confirm that Enron provides a conventional organizational communication environment against which the more unusual structure of Jmail can later be interpreted.

\vspace{\compactBeforeSubsection}
\subsubsection{Ego network structure.}
\begin{figure*}[!t]
    \centering
    \begin{subfigure}[!h]{0.48\textwidth}
        \centering
        \includegraphics[trim=0 0 0 5cm,clip,width=\linewidth]{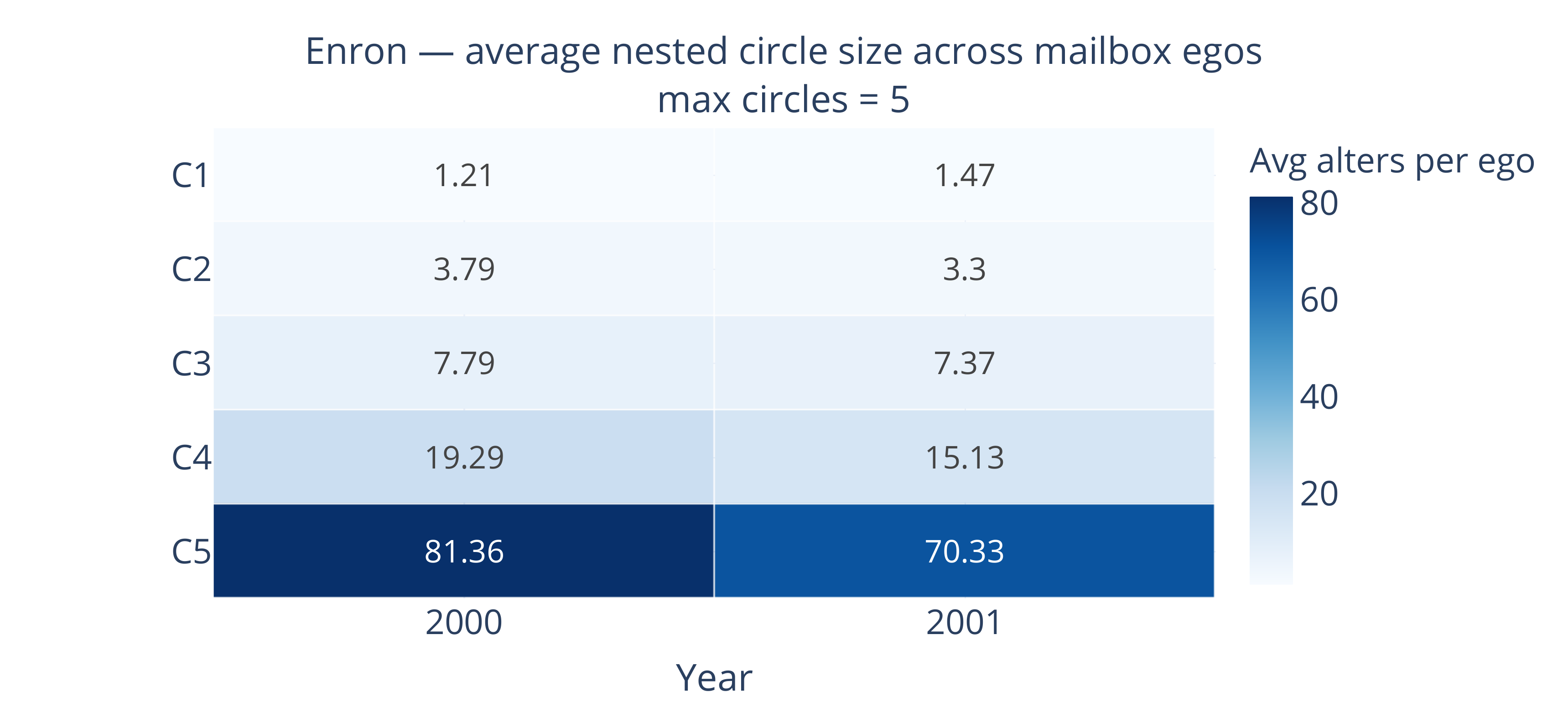}
        \caption{Circle sizes across years}
        \label{fig:enron_dynamic_sizes}
    \end{subfigure}
    \hfill
     \begin{subfigure}[!h]{0.48\textwidth}
        \centering
        \includegraphics[trim=0 0 0 5cm,clip,width=\linewidth]{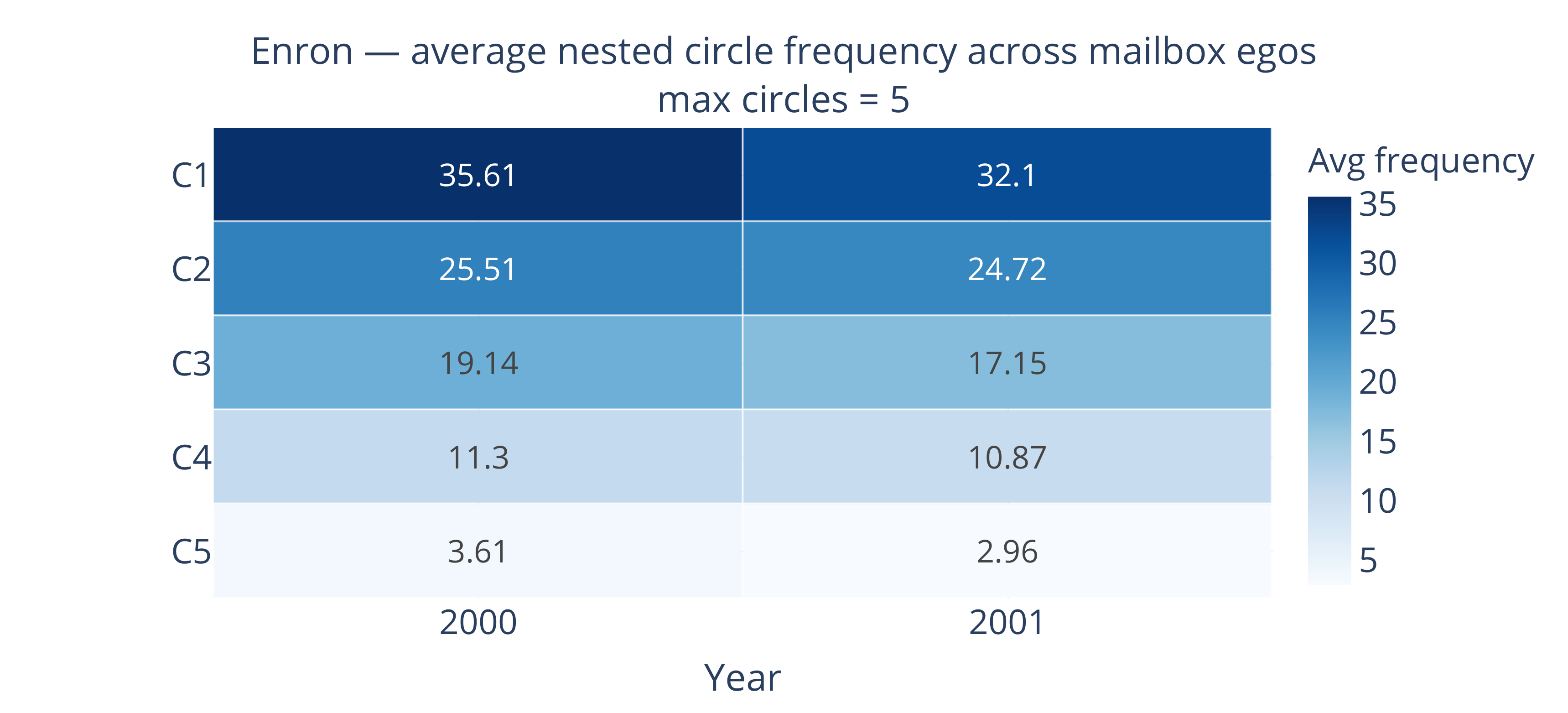}
        \caption{Average annual frequency by circle}
        \label{fig:enron_dynamic_freq}
    \end{subfigure}
    \caption{Enron --- Five-circle ego network structure in the most active years}
    \label{fig:enron_dynamic_summary}
    \vspace{\compactAfterFigure}
\end{figure*}

We then narrow the analysis to the most active years, where ego networks are more likely to be sufficiently populated and structurally informative. Year 2001 counts 156 active egos, while the second most populated year is 2000 with 127. We thus focus on 2001 for our comparisons with Jmail. Among the active Enron ego networks in 2001, the most common configurations contain between four and seven circles: 24 egos exhibit four circles, 30 exhibit five circles, 22 exhibit six circles, and 25 exhibit seven circles, with five-circle structures being the most frequent case. We therefore focus on the five-circle egos in 2001.
When looking at the size and interaction frequency for their circles, Figure~\ref{fig:enron_dynamic_summary} shows that Enron partially reproduces the layered organization predicted by Dunbar's model. Cumulative circle sizes close to $\sim$5, $\sim$15, and $\sim$50 alters emerge consistently across years, while the broader $\sim$150 layer is not recovered. Instead, an additional innermost layer of approximately $\sim$3 alters appears, consistent with prior observations of digitally mediated ego networks.
In frequency terms, the strongest correspondence with the Dunbar reference appears in the outermost observed layer. Although this layer has the size of the Dunbar $\sim$50 circle rather than the $\sim$150 circle, its average contact frequency is close to the corresponding semiannual threshold: approximately 3 messages per year, compared with 2 in the Dunbar model. The inner circles recover Dunbar-like sizes but at lower-than-expected frequencies, suggesting that organizational email captures the structural skeleton of personal networks but compresses the frequency range relative to offline interaction.

\begin{tcolorbox}[keyfindingbox, title={Key Finding 1: Workplace email recovers partial Dunbar-like structure}]
Enron produces stable ego network layers that broadly resemble Dunbar-style organization. Small inner circles and an outer $\sim$50 layer emerge consistently, but the broader $\sim$150 layer is absent, suggesting that workplace email primarily captures recurrent organizational ties rather than complete personal social networks.
\end{tcolorbox}

\subsection{Breakdown and Recovery of Layered Structure in Jmail}
\label{sec:results_jmail}
\vspace{-4pt}

We now apply the same analysis to Jmail, the focal archive for evaluating Dunbar-like ego network organization in a high-frequency ego-centered setting. The Enron benchmark established that Dunbar-like layering is recoverable from workplace email; the question here is whether it remains recoverable when communication is unusually intense. We proceed in four steps: temporal activity, empirical ego network layers, Dunbar-reference assignment and scaling, and reciprocity.


\vspace{\compactBeforeSubsection}
\subsubsection{Temporal activity patterns.} 
\begin{figure}[!t]
    \centering

    \begin{subfigure}[t]{1\linewidth}
        \centering
        \includegraphics[trim=0 0 0 0.9cm,clip,width=\linewidth]{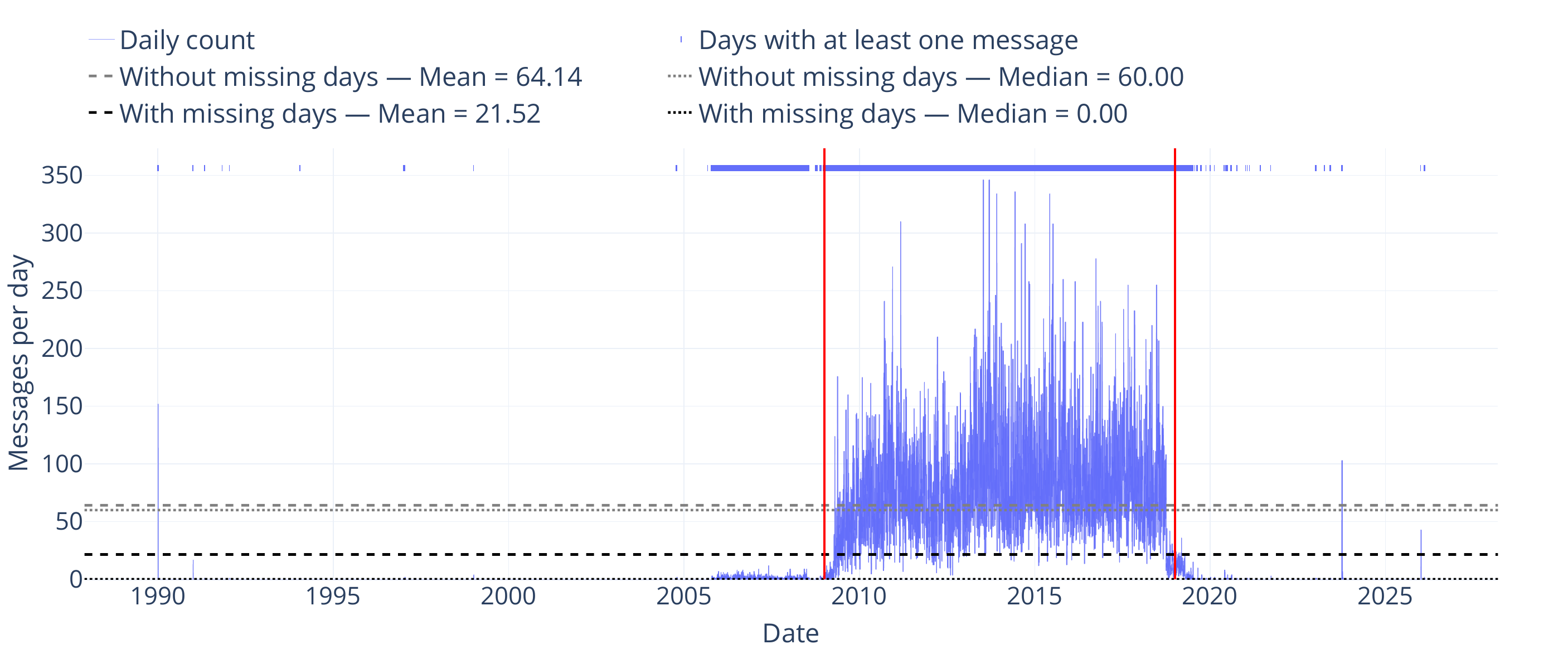}
        \caption{Daily activity}
        \label{fig:jeff_daily}
    \end{subfigure}


    \begin{subfigure}[t]{0.45\linewidth}
        \centering
        \includegraphics[trim=0 0 0 2.5cm,clip,width=\linewidth]{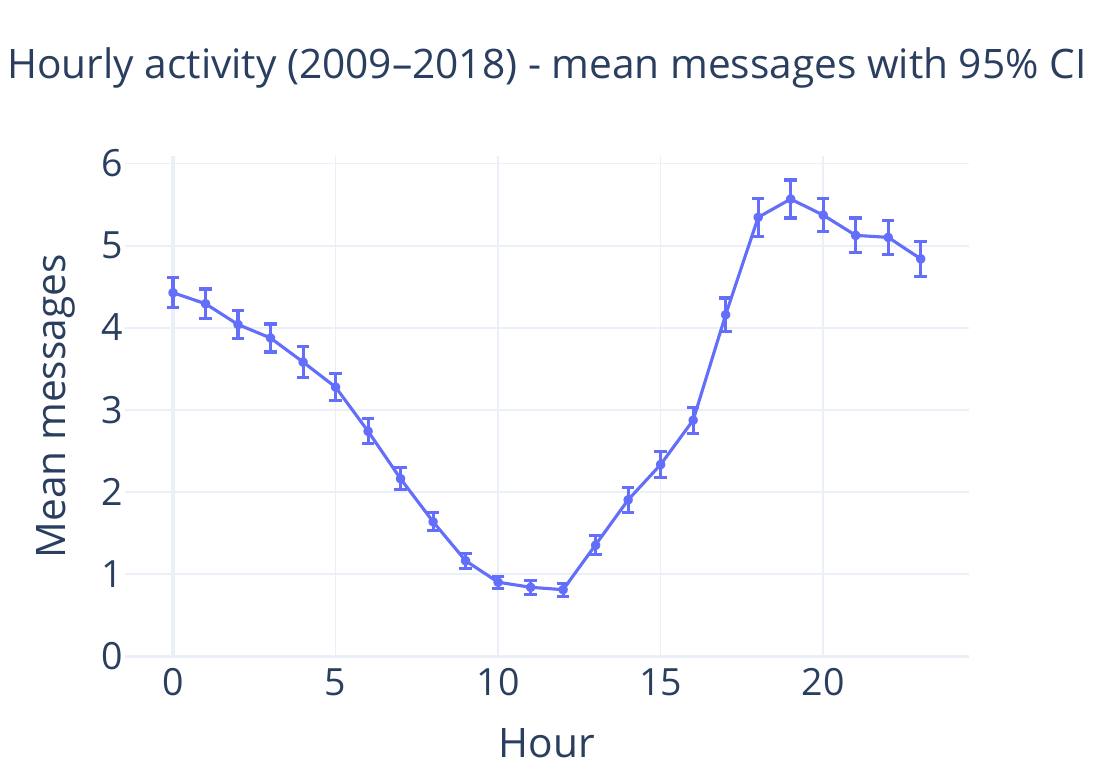}
        \caption{Hourly activity}
        \label{fig:jeff_hourly}
    \end{subfigure}
    \hfill
    \begin{subfigure}[t]{0.45\linewidth}
        \centering
        \includegraphics[trim=0 0 0 2.5cm,clip,width=\linewidth]{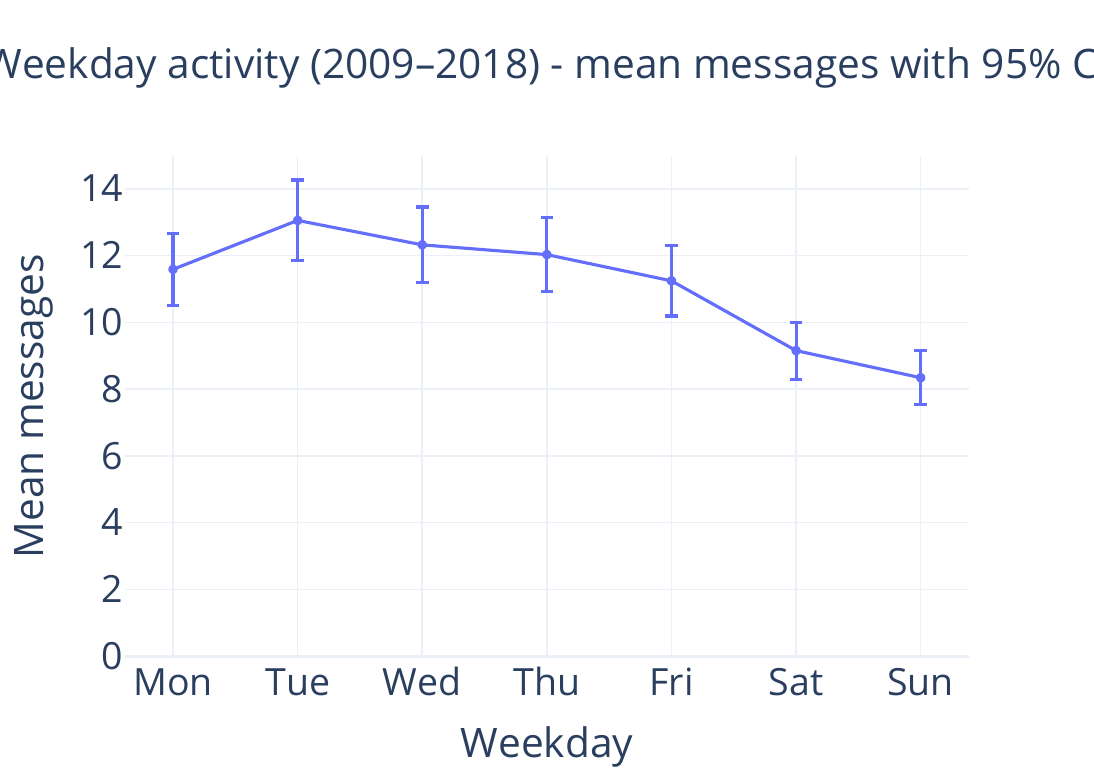}
        \caption{Weekday activity}
        \label{fig:jeff_weekly}
    \end{subfigure}

    \caption{Jmail (2009-2018) --- Temporal activity patterns of the main ego. }
    \label{fig:jeff_temporal_activity}
    \vspace{\compactAfterFigure}
\end{figure}

Figure~\ref{fig:jeff_temporal_activity} reports daily, hourly, and weekday activity for the Jmail main ego. The core period spans 2009--2018, with an average of approximately 64 messages per active day---roughly one hundred times the Enron average. Unlike Enron, hourly activity is lowest during standard working hours and rises sharply after 16:00, peaking around 19:00--20:00, with substantial activity also at night and in the early morning. The weekday profile is broadly similar to Enron, with higher activity early in the week and a gradual decline toward the weekend, though activity remains high across all days. Together, these patterns confirm that Jmail is not a workplace archive: its temporal rhythm reflects personal rather than organizational communication.

\vspace{\compactBeforeSubsection}
\subsubsection{Empirical ego network layers.}
\vspace{\compactAfterSubsection}

\begin{figure*}[!b]
\vspace{\compactBeforeFigure}
    \centering

    \begin{subfigure}[!h]{0.49\textwidth}
        \centering
        \includegraphics[trim=0 0 0 2.5cm,clip,width=1.06\linewidth]{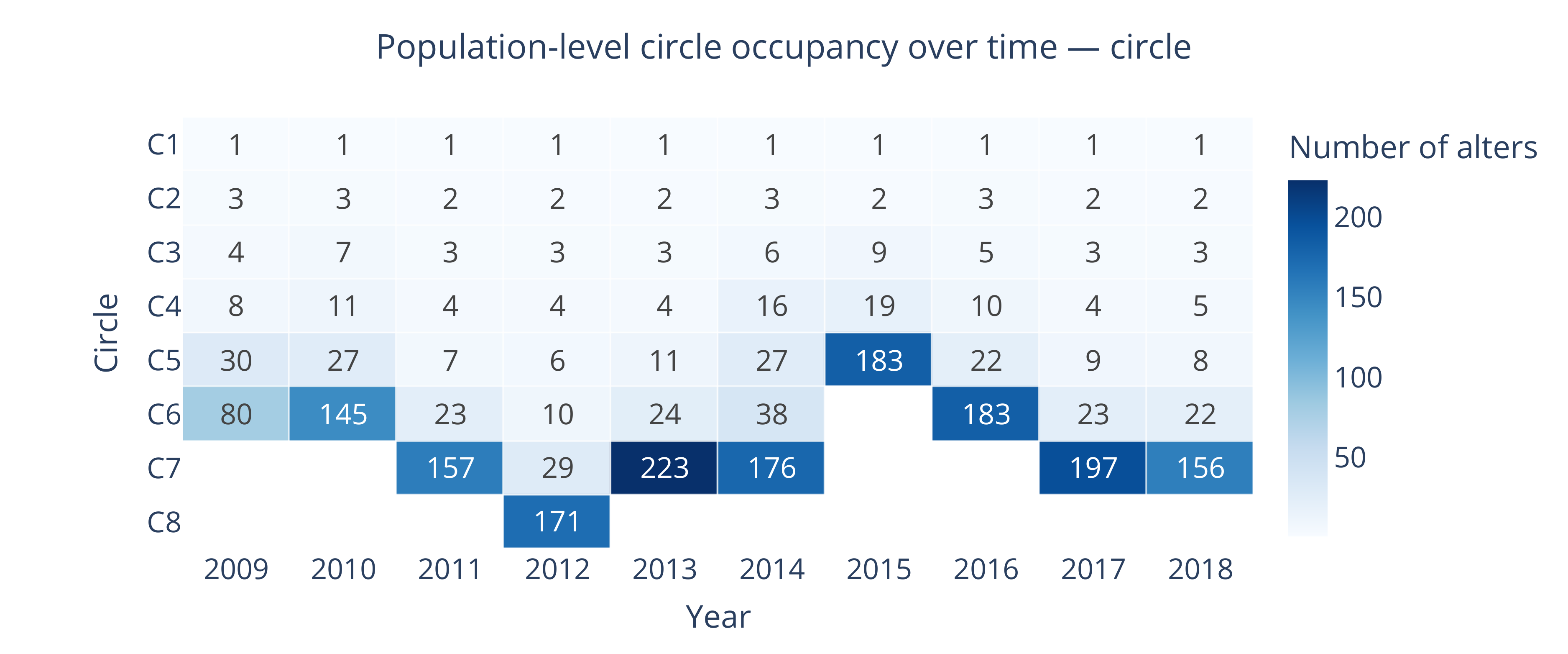}
        \caption{Circle sizes across years}
        \label{fig:jeff_dynamic_sizes}
    \end{subfigure}
    \hfill
    \begin{subfigure}[!h]{0.49\textwidth}
        \centering
        \includegraphics[trim=0 0 0 2.5cm,clip,width=1.1\linewidth]{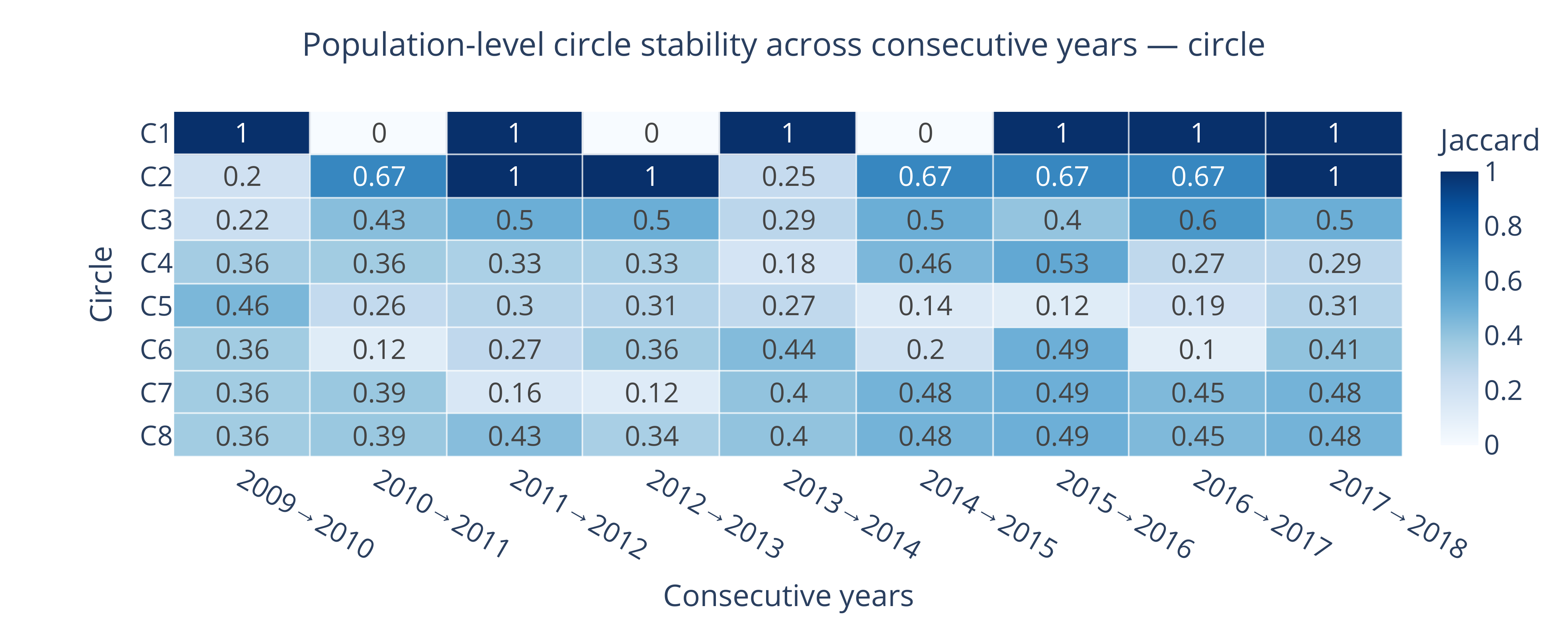}
        \caption{Stability of circle membership across consecutive years}
        \label{fig:jeff_dynamic_jaccard}
    \end{subfigure}
    \hfill
     \begin{subfigure}[!h]{0.7\textwidth}
        \centering
        \includegraphics[trim=0 0 0 2.5cm,clip,width=\linewidth]{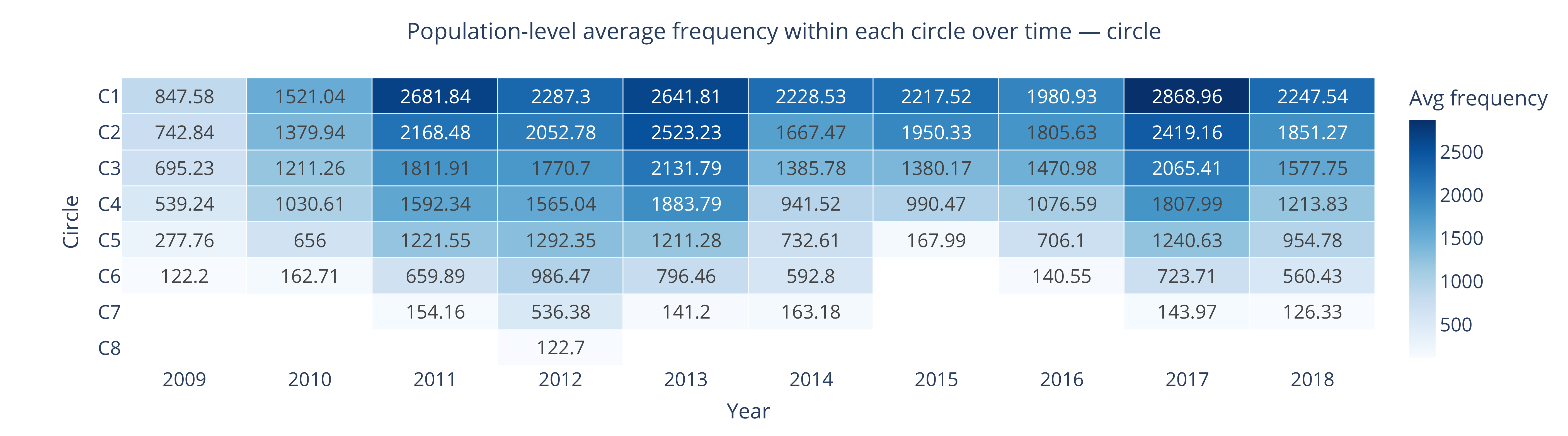}
        \caption{Average annual frequency by circle}
        \label{fig:jeff_dynamic_freq}
    \end{subfigure}
    \caption{Jmail --- Empirical MeanShift circle structure of the main ego.}
    \label{fig:jeff_dynamic_summary}
    \vspace{\compactAfterFigure}

\end{figure*}

With this context in mind, we ask whether the higher communication intensity observed in Jmail produces stable ego network layers. Figure~\ref{fig:jeff_dynamic_summary}(a) summarizes the empirical MeanShift circles extracted from annual ego--alter interaction frequencies. Two features stand out. First, as in Enron, an additional inner layer of approximately $\sim$3 alters emerges. Second, unlike Enron, Jmail reaches the broader $\sim$150-alter scale, consistent with a more complete personal network. However, the empirical structure is highly unbalanced: inner circles are very small while the outermost circle concentrates the majority of active alters, and this imbalance is not simply a reflection of a naturally skewed network.

Figure~\ref{fig:jeff_dynamic_summary}(b) shows that circle membership is measurable across consecutive years, with Jaccard stability generally higher in the inner circles than in the outer ones. This is consistent with the frequency imbalance: the innermost circles contain the ego's most frequently and persistently contacted alters, whose high interaction rates make them less likely to migrate across layers from year to year. The outer circles, by contrast, aggregate a larger and more heterogeneous pool of contacts whose annual frequencies are more variable, resulting in lower membership stability. This pattern suggests that despite the structural imbalance observed in circle sizes, the high-frequency core is not an artifact of year-to-year fluctuations but reflects a stable set of privileged relationships.

The frequency profile in Figure~\ref{fig:jeff_dynamic_summary}(c) reveals the cause of the structural imbalance in Figure~\ref{fig:jeff_dynamic_summary}(a). The innermost empirical circles reach annual frequencies close to 2,000 messages---far above both the Dunbar weekly-contact reference of 53 per year and the Enron inner-circle value of approximately 35. Even the outermost empirical layer exceeds 100 messages per year, which is itself above the Dunbar threshold for the most active layer. Jmail therefore contains a few extremely frequent contacts and a much larger group of alters that are less intense only in relative terms, but still highly active in absolute terms. This extreme skew causes MeanShift to fragment the high-frequency core into very sparse inner circles while placing many still-active contacts in the broad outer layer. The empirical layering is therefore hard to interpret in Dunbar terms, which motivates a more principled diagnostic approach.




\vspace{\compactBeforeSubsection}
\subsubsection{Dunbar-reference circles and scaling}
\vspace{\compactAfterSubsection}


\begin{figure*}[!t]

    \centering

    \begin{subfigure}[t]{0.48\textwidth}
        \centering
        \includegraphics[trim=0 0 0 2.5cm,clip,width=1.1\linewidth]{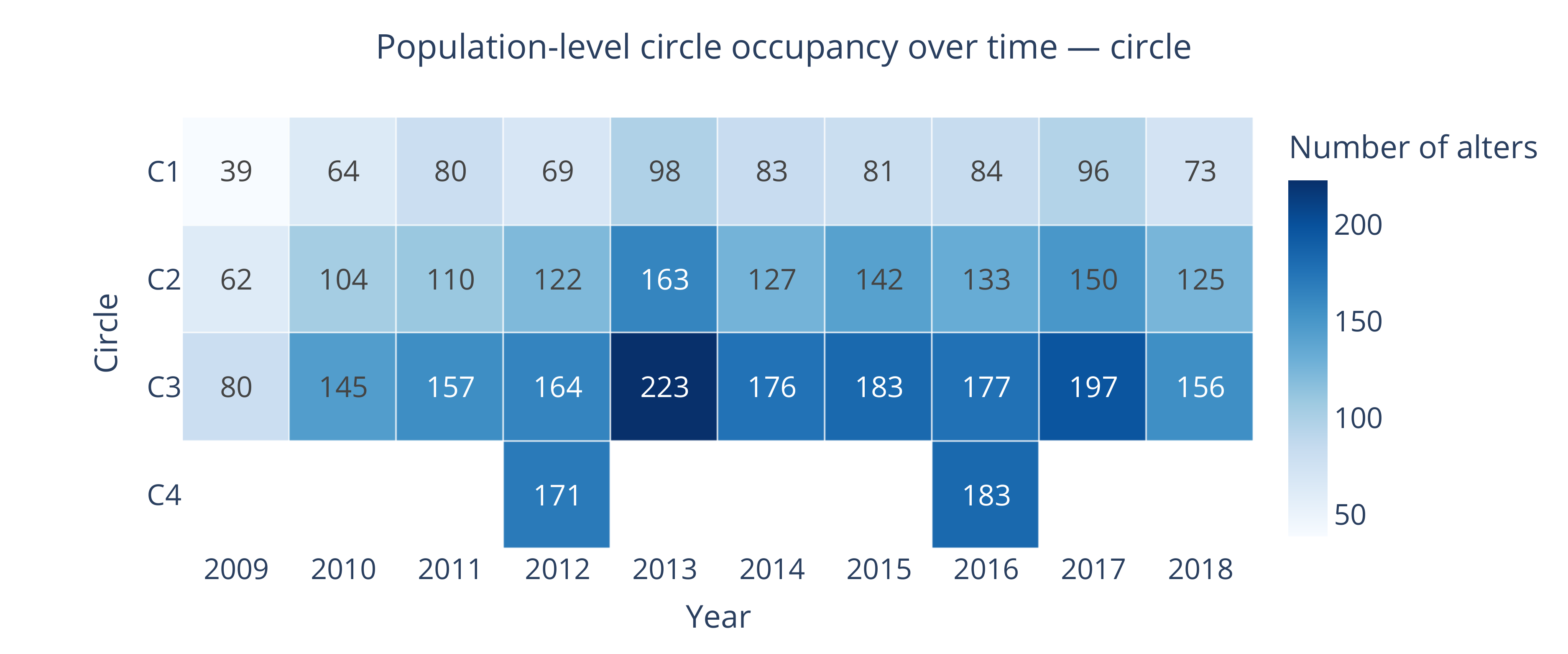}
        \caption{Circle sizes across years}
        \label{fig:jeff_dynamic_sizes_dunbar}
    \end{subfigure}
    \hfill
    \begin{subfigure}[t]{0.48\textwidth}
        \centering
        \includegraphics[trim=0 0 0 2.5cm,clip,width=1.1\linewidth]{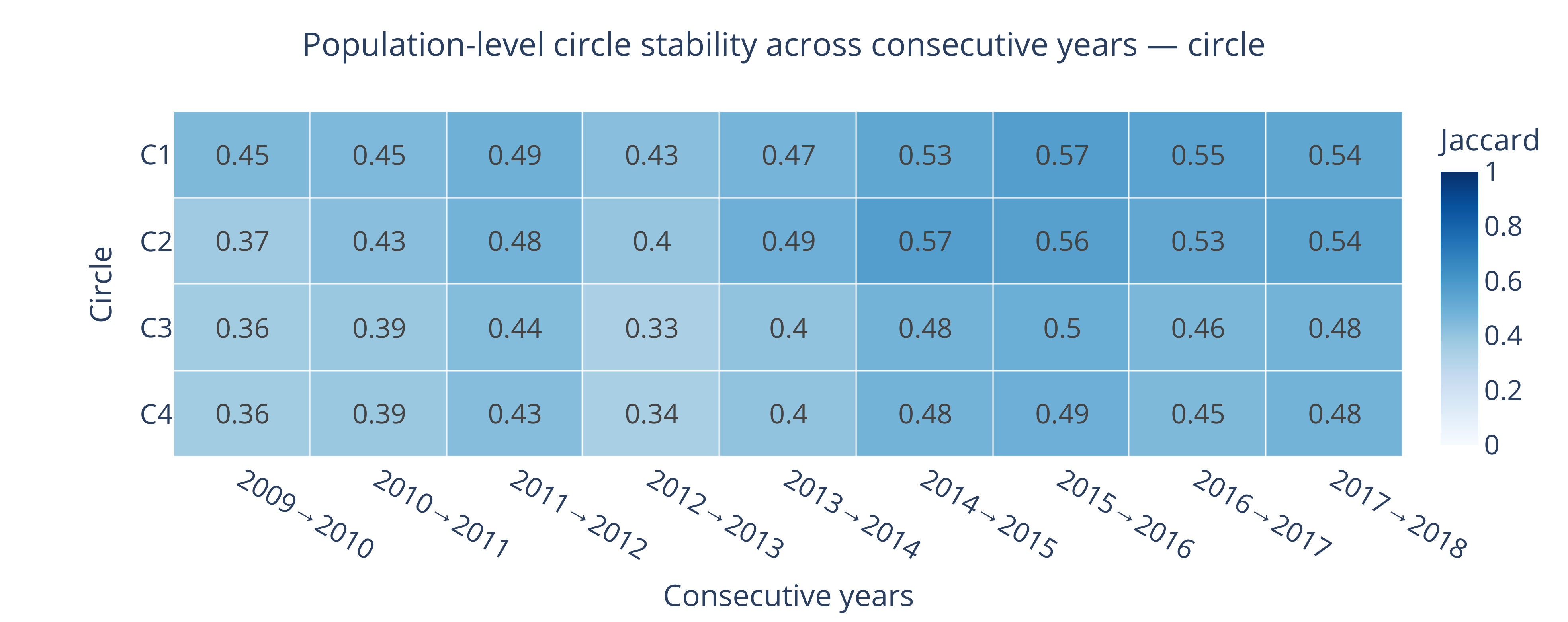}
        \caption{Stability of circle membership across consecutive years}
        \label{fig:jeff_dynamic_jaccard_dunbar}
    \end{subfigure}


    \begin{subfigure}[t]{0.6\textwidth}
        \centering
        \includegraphics[trim=0 0 0 2.5cm,clip,width=\linewidth]{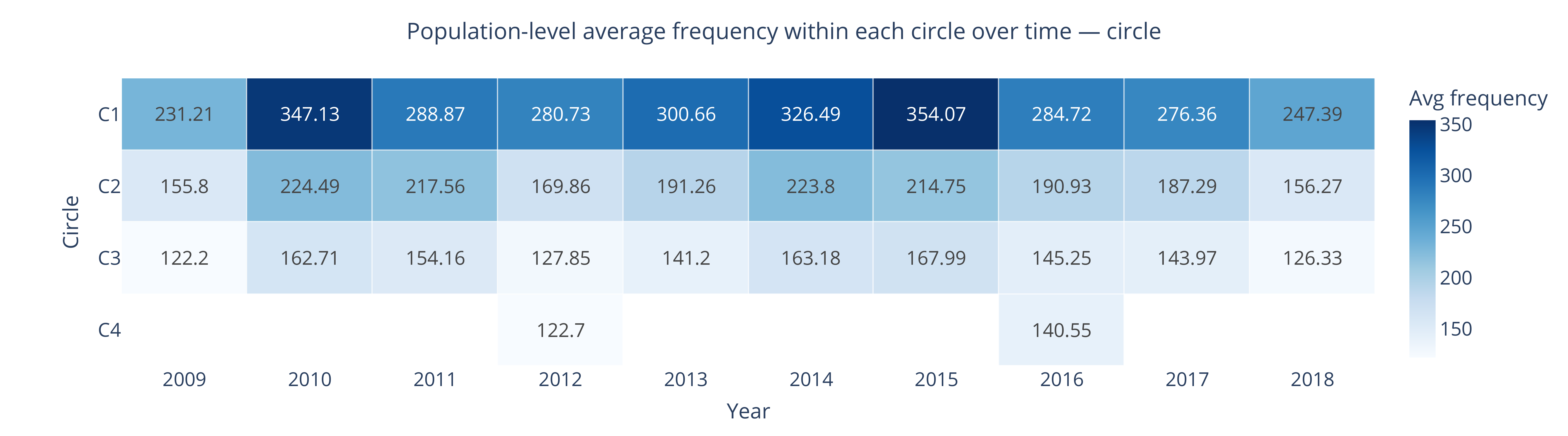}
        \caption{Average annual frequency by circle}
        \label{fig:jeff_dynamic_freq_dunbar}
    \end{subfigure}

    \caption{Jmail --- Dunbar-reference circle structure under canonical frequency thresholds.}
    \label{fig:jeff_dynamic_summary_dunbar}
    \vspace{\compactAfterFigure}

\end{figure*}

To diagnose this distortion more precisely, we apply the canonical Dunbar frequency thresholds directly to Jmail as a baseline. Specifically, we ignore the circles created by MeanShift and assign alters to circles based on the canonical Dunbar frequency thresholds. E.g., with reference to Table~\ref{tab:egonet_offline}, a frequency higher than once a week would put the alter in circle~1. 

Figure~\ref{fig:jeff_dynamic_summary_dunbar} shows that the result is not simply a noisy version of an ordinary ego network: the unscaled assignment produces an almost reversed profile, with many alters falling into inner or intermediate Dunbar circles and the outermost yearly-contact layer weakly represented. This reversal does not indicate an absence of structure. Rather, it confirms that Jmail is a selected pool of high-interest alters whose baseline interaction frequencies are already high relative to the canonical Dunbar scale, so the standard thresholds are not appropriate as-is. The solution is not to abandon Dunbar's model, but to adapt its frequency scale to Jmail's communication regime while preserving its structural logic.

We therefore rescale the Dunbar frequency ladder while preserving the canonical circle sizes and their ordering. The rescaling is anchored to one of Dunbar's circles: for a given reference circle of size $N$, we rank the ego's active alters by annual frequency in each year, take the frequency of the $N$th-ranked alter, and average these values across the observation period. This empirical frequency is then divided by the Dunbar contact threshold associated with that circle, yielding a scaling factor that re-expresses the full Dunbar ladder in Jmail's frequency units without forcing any particular layered structure to appear. We use the support clique ($N=5$, weekly contact) as the reference circle. We anchor the scaling procedure to the support clique because it is the most stable and cognitively constrained layer in Dunbar’s model. As the ego’s strongest and most regularly maintained relationships, these ties are less affected by contextual and platform-specific variability than broader outer layers. In Jmail, the support clique also provides the clearest empirical boundary in an otherwise highly skewed communication environment, making it a robust reference for adapting Dunbar’s frequency scale to the archive’s unusually high interaction intensity. The resulting scaling factor using the support clique is 17. A sensitivity check anchored to the sympathy group ($N=15$, monthly contact) is reported in \appref{app:jmail_scaled} and yields consistent results.

\begin{figure}[!t]
    \centering

    \begin{subfigure}[!h]{0.48\textwidth}
        \centering
        \includegraphics[trim=0 0 0 2.5cm,clip,width=1.1\linewidth]{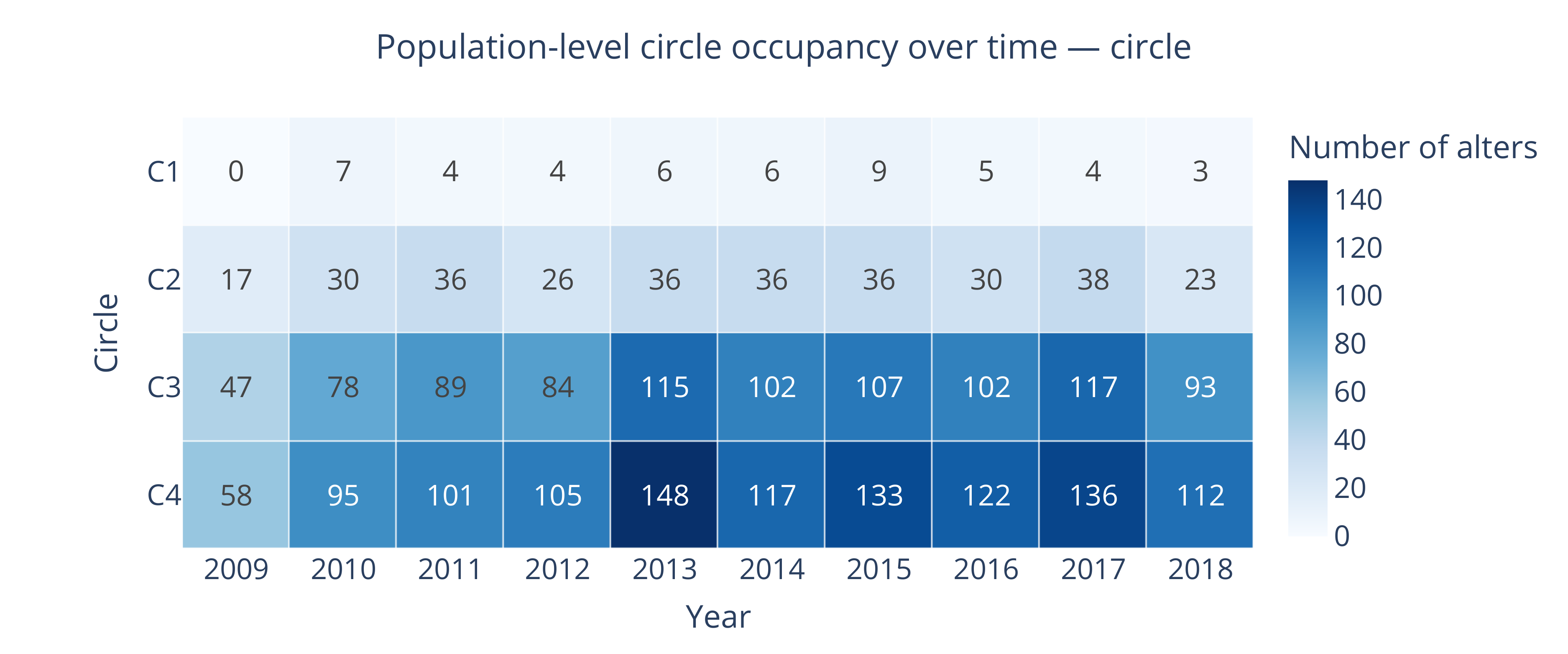}
        \caption{Circle sizes across years}
        \label{fig:jeff_dynamic_sizes_dunbar-top5}
    \end{subfigure}
    \hfill
    \begin{subfigure}[!h]{0.48\textwidth}
        \centering
        \includegraphics[trim=0 0 0 2.5cm,clip,width=1.1\linewidth]{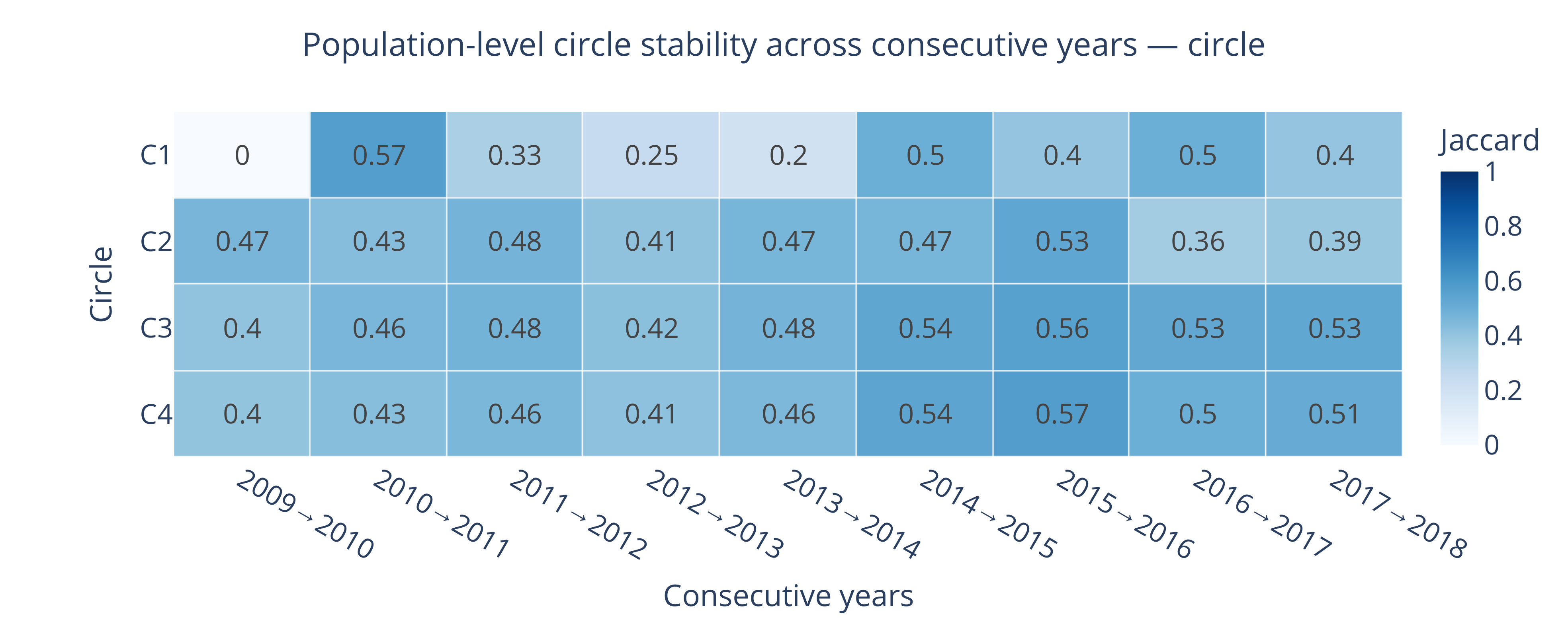}
        \caption{Stability (Jaccard) of circle membership across consecutive years}
        \label{fig:jeff_dynamic_jaccard_dunbar-top5}
    \end{subfigure}
    \hfill
     \begin{subfigure}[!h]{0.8\textwidth}
        \centering
        \includegraphics[trim=0 0 0 2.5cm,clip,width=0.8\linewidth]{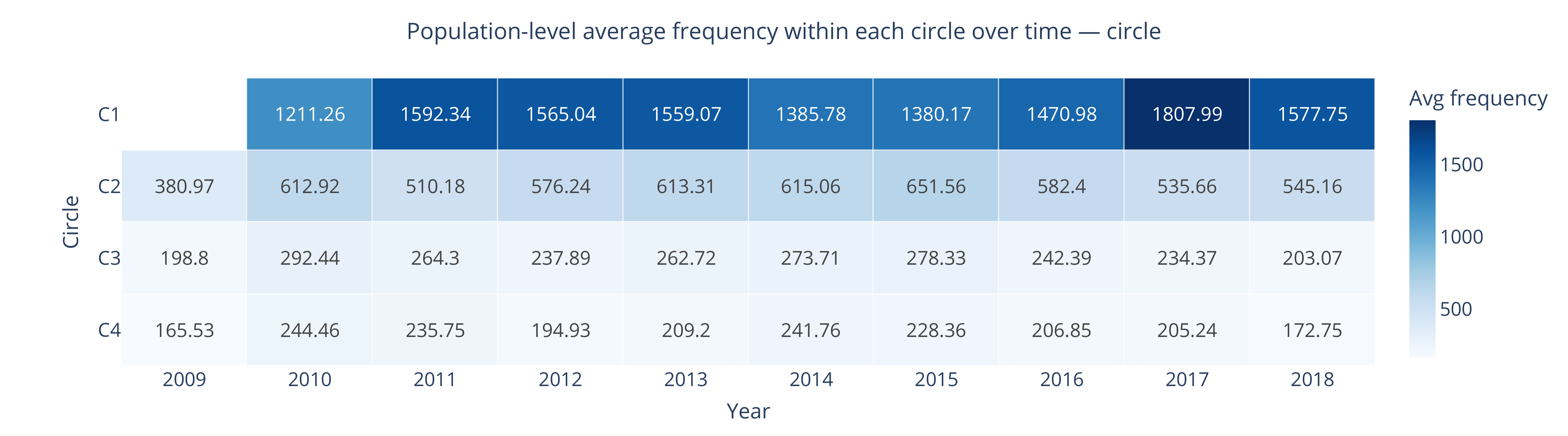}
        \caption{Average frequency of circle membership across consecutive years}
        \label{fig:jeff_dynamic_freq_dunbar-top5}
    \end{subfigure}
    \caption{Jmail --- Top-$5$ scaled Dunbar-reference circles.}
    \label{fig:jeff_dynamic_summary_dunbar_top5}
\vspace{-0.5cm}
\end{figure}

Figure~\ref{fig:jeff_dynamic_summary_dunbar_top5} shows that, once the Dunbar ladder is anchored to the observed support-clique boundary, a clearer layered structure emerges. The rescaled circles recover a more interpretable Dunbar-like organization in size, preserve measurable temporal stability, and retain a decreasing frequency gradient from the high-contact core toward broader layers. Importantly, this structure is not an artifact of the scaling procedure: because scaling only changes the frequency unit of the reference thresholds, the emergence of coherent layers is not guaranteed by construction. This result shows that Jmail does contain Dunbar-like layered organization, but that it is obscured when standard methods are applied without accounting for the archive's unusually high communication intensity. A natural remaining question is whether these recovered layers reflect genuine social relationships or are simply an artifact of the main ego's outgoing activity.

\begin{tcolorbox}[keyfindingbox, title={Key Finding 2: Jmail requires scaled Dunbar-reference thresholds}]
Jmail does not behave like a complete ego network with many weak peripheral contacts, but like a selected pool of high-interest alters. This extreme frequency regime distorts both data-driven clustering and unscaled Dunbar thresholds. Once Dunbar frequencies are scaled to the activity of the high-contact core, the main layered structure re-emerges, showing that Dunbar's model remains useful as a scalable structural reference.
\end{tcolorbox}

\vspace{\compactBeforeSubsection}
\subsubsection{Reciprocity of scaled ego network layers.} 
The scaled layers reorganize Jmail's alters into a more interpretable structure, but a residual concern remains: are these layers a genuine reflection of social relationships, or simply an artifact of the main ego's outgoing activity? To address this, for each year we retain alters for whom communication is observed in both directions (main ego $\rightarrow$ alter and alter $\rightarrow$ main ego) and test whether alters placed in higher-frequency Dunbar circles also contact the main ego more frequently in their own outgoing communication. Table~\ref{tab:jmail_reciprocity_top5_last3} reports the results for the last three years; full year-wise tables for all scaling variants are reported in 
\appref{app:jmail_reciprocity_sensitivity}.
The overall ego--alter frequency correlations are consistently and strongly positive, increasing from $r=0.91$ in 2016 to $r=0.95$ in 2018, confirming that the scaled layers capture genuine bidirectional relationships rather than one-sided ego activity. Circle-wise correlations are more variable, particularly in the innermost circle where few reciprocal alters are available. The middle and outer circles are more informative: Circle~3 contains 14, 14, and 10 reciprocal alters in 2016--2018 with positive correlations throughout, and Circle~4 remains positively correlated in the same period. Finally, the MeanShift composition column confirms that the scaled Dunbar circles consolidate the over-fragmented empirical core into broader, more coherent layers, whereas the unscaled assignment collapses many heterogeneous data-driven rings into a single Dunbar circle.

\begin{tcolorbox}[keyfindingbox, title={Key Finding 3: Scaled Jmail layers are reciprocally meaningful}]
The scaled layers reflect genuine bidirectional relationships: alters in higher-frequency circles also contact the main ego more actively, with overall correlations consistently above $r=0.91$. Frequency scaling thus recovers both structural interpretability and relational meaningfulness.
\end{tcolorbox}

\vspace{\compactBeforeTable}
\begin{table*}[!h]
\centering
\tiny
\setlength{\tabcolsep}{4pt}
\renewcommand{\arraystretch}{1.15}
\caption{Jmail — Reciprocity under scaled Dunbar-reference circles. The table includes alters who had at least one contact per year with the main actor.}
\label{tab:jmail_reciprocity_top5_last3}
\resizebox{\linewidth}{!}{
\begin{tabular}{lcccccccccc}
\toprule
Year & \multicolumn{2}{c}{\cellcolor{gray!10}Overall} & \multicolumn{2}{c}{\cellcolor{red!8}Circle 1} & \multicolumn{2}{c}{\cellcolor{yellow!14}Circle 2} & \multicolumn{2}{c}{\cellcolor{green!10}Circle 3} & \multicolumn{2}{c}{\cellcolor{blue!8}Circle 4} \\
\cmidrule(lr){2-3}\cmidrule(lr){4-5}\cmidrule(lr){6-7}\cmidrule(lr){8-9}\cmidrule(lr){10-11}
& \cellcolor{gray!10}$n$ & \cellcolor{gray!10}$r$ & \cellcolor{red!8}$n$ & \cellcolor{red!8}$r$ & \cellcolor{yellow!14}$n$ & \cellcolor{yellow!14}$r$ & \cellcolor{green!10}$n$ & \cellcolor{green!10}$r$ & \cellcolor{blue!8}$n$ & \cellcolor{blue!8}$r$ \\
\midrule
2016 & \cellcolor{gray!10}37 & \cellcolor{gray!10}0.91 & \cellcolor{red!8}3 & \cellcolor{red!8}-0.11 & \cellcolor{yellow!14}6 & \cellcolor{yellow!14}0.21 & \cellcolor{green!10}14 & \cellcolor{green!10}0.62 & \cellcolor{blue!8}7 & \cellcolor{blue!8}0.55 \\
2017 & \cellcolor{gray!10}34 & \cellcolor{gray!10}0.92 & \cellcolor{red!8}3 & \cellcolor{red!8}0.34 & \cellcolor{yellow!14}6 & \cellcolor{yellow!14}0.75 & \cellcolor{green!10}14 & \cellcolor{green!10}0.89 & \cellcolor{blue!8}4 & \cellcolor{blue!8}0.94 \\
2018 & \cellcolor{gray!10}33 & \cellcolor{gray!10}0.95 & \cellcolor{red!8}2 & \cellcolor{red!8}-1.00 & \cellcolor{yellow!14}6 & \cellcolor{yellow!14}0.34 & \cellcolor{green!10}10 & \cellcolor{green!10}0.40 & \cellcolor{blue!8}5 & \cellcolor{blue!8}0.86 \\
\bottomrule
\end{tabular}
}
\end{table*}
\vspace{\compactAfterTable}

\vspace{-10pt}
\section{Conclusion}
\vspace{\compactAfterSection}

\label{sec:conclusion}

This paper examined whether layered ego network structure can be recovered from email communication. Using Enron as a workplace benchmark, we showed that email-derived ego networks can reproduce part of the Dunbar-like organization observed in other communication media: small inner circles are visible, and the outermost observed layer is close to the Dunbar $\sim$50 circle, while the broader $\sim$150 circle is not recovered.
We then analysed Jmail as a contrasting, high-frequency ego-centered archive. The case-specific nature of Jmail is therefore part of the empirical contribution: it provides a rare public setting in which layered ego network methods can be tested on a highly ego-centered, socially salient email archive. The results show that Jmail should not be interpreted as an ordinary email ego network with higher activity, but as a selected pool of high-interest alters where weak and occasional contacts are comparatively underrepresented. This affects both data-driven and reference-based approaches: MeanShift separates a few extremely frequent alters into narrow inner circles, while unscaled Dunbar thresholds produce an almost reversed ego network profile. When the Dunbar frequency scale is adapted to the high-contact core, a clearer layered structure emerges.
Finally, reciprocity analysis suggests that the reconstructed layers are not only artifacts of the main ego's activity. Reciprocal active alters show positive ego--alter frequency correlations, and the scaled circles provide a smoother interpretation of the empirical MeanShift structure. Overall, the findings suggest that Dunbar's model remains useful for email ego networks. Enron shows that Dunbar-like layering can emerge in ordinary workplace email, while Jmail shows that selective high-frequency archives require normalization before the same structure becomes interpretable. Future work should extend this analysis to additional email archives and examine whether similar scaling behavior appears in other high-intensity communication environments.

\vspace{-10pt}



%
%
%
\bibliographystyle{splncs04}
\bibliography{bibliography}

\begin{thebibliography}{8}
\bibitem{ref_article1}
Author, F.: Article title. Journal \textbf{2}(5), 99--110 (2016)

\bibitem{ref_lncs1}
Author, F., Author, S.: Title of a proceedings paper. In: Editor,
F., Editor, S. (eds.) CONFERENCE 2016, LNCS, vol. 9999, pp. 1--13.
Springer, Heidelberg (2016). \doi{10.10007/1234567890}

\bibitem{ref_book1}
Author, F., Author, S., Author, T.: Book title. 2nd edn. Publisher,
Location (1999)

\bibitem{ref_proc1}
Author, A.-B.: Contribution title. In: 9th International Proceedings
on Proceedings, pp. 1--2. Publisher, Location (2010)

\bibitem{ref_url1}
LNCS Homepage, \url{http://www.springer.com/lncs}, last accessed 2023/10/25
\end{thebibliography}
%

\end{document}